\documentclass[journal]{IEEEtran}
\ifCLASSINFOpdf
\else
\fi
\usepackage{comment}
\usepackage{bm}
\usepackage{amssymb}
\usepackage{enumitem}
\usepackage{amssymb}
\usepackage{mathptmx}
\usepackage{url}
\usepackage[scaled=.90]{helvet}
\usepackage{courier}
\usepackage{caption}
\usepackage{subcaption}
\usepackage{datetime}
\usepackage{nomencl} 
\usepackage{hyperref}
\usepackage[Sonny]{fncychap} 
\usepackage{makeidx}
\usepackage{float}
\usepackage{amsmath}
\usepackage{amssymb}

\usepackage[figuresright]{rotating}
\usepackage{xcolor}
\usepackage{multirow}

\begin{document}
%

\title{Relay-based identification of Aerodynamic and Delay Sensor Dynamics with applications for Unmanned Aerial Vehicles }

%
%
%

\author{Anees~Peringal,
        Mohamad~Chehadeh,~\IEEEmembership{Member, IEEE,}
        and~Igor~Boiko,~\IEEEmembership{Senior Member,~IEEE}
        and~Yahya~Zweiri,~\IEEEmembership{Member, IEEE}
\thanks{This work was supported by the Khalifa University of Science and Technology under Award CIRA-2020-082 and Award RCI-2018-KUCARS.}
\thanks{A. Peringal, M. Chehadeh and Y. Zweiri are with the Department of Aerospace Engineering, Khalifa University, Abu Dhabi, United Arab Emirates and Khalifa University Center for Autonomous Robotic Systems (KUCARS), Khalifa University, Abu Dhabi, United Arab Emirates.}
\thanks{Y. Zweiri is the Director of the Advanced Research and Innovation Center (ARIC), Khalifa University,
Abu Dhabi, United Arab Emirates.}
\thanks{I. BOIKO is with the Department of Electrical Engineering and Computer Science, Khalifa University, Abu Dhabi, United Arab Emirates.}}

%
%

\markboth{Journal of \LaTeX\ Class Files,~Vol.~14, No.~8, August~2015}%
{Shell \MakeLowercase{\textit{et al.}}: Bare Demo of IEEEtran.cls for IEEE Journals}
%



\maketitle

\begin{abstract}
In this paper, we present a real-time system identification method based on relay feedback testing with applications to multirotor unmanned aerial vehicles. The proposed identification method provides an alternative to the expensive lab testing of certain UAV dynamic parameters. Moreover, it has the advantage of identifying the parameters that get changed throughout the operation of the UAV which requires onboard identification methods. The modified relay feedback test (MRFT) is used to generate stable limit cycles at frequency points that reveal the underlying UAV dynamics. The locus of the perturbed relay system (LPRS) is used to predict the exact amplitude and frequency of these limit cycles. Real-time identification is achieved by using the homogeneity properties of the MRFT and the LPRS which are proven in this paper. The proposed identification method was tested experimentally to estimate the aerodynamic parameters as well as the onboard sensor's time delay parameters. The MRFT testing takes a few seconds to perform, and the identification computations take an average of 0.2 seconds to complete in modern embedded computers. The proposed identification method is compared against state-of-the-art alternatives. Advantages in identification accuracy and quantification of uncertainty in estimated parameters are shown.
\end{abstract}

\begin{IEEEkeywords}
Sensor signal processing, robotics and automation applications, sensor decision and fusion.
\end{IEEEkeywords}

%
\IEEEpeerreviewmaketitle

\section{Introduction}
\label{introduction}
\subsection{Literature Review}
System identification is the process of obtaining the mathematical model of a dynamical system by observing the system response to a certain input excitation \cite{ljung2010perspectives}. Obtaining a model is equivalent to building a hypothesis about the system, which can be used to conveniently analyze system properties. System identification requires a model structure, data observations to fit the model, and a set of rules to fit the data to the model structure, which is done with the help of a metric to quantify the quality of the developed model. Since the possible specifications on models, data, and metrics are quite vast, the literature offers a plethora of system identification methods, each with its own pros and cons.

System identification techniques may be categorized into grey-box or black-box methods based on the model structure considered. A model is referred to as a grey-box model when it is based on physical principles. Some examples of grey-box models are the steel frame structure like the one used in \cite{Hann2009Sensors}, and the Wiener model used in the identification of creep in \cite{Qi2021Sensors}. On the other hand, black-box models can be arbitrarily complex, and are thus more versatile but are hardly usable for subsequent design, e.g. controller or estimator design. Some examples of black-box models for system identification would use different classes of neural networks to learn the dynamics of the system based on its input-output data. In \cite{Jiang2022}, a feed-forward neural network is used to learn the dynamics of a UAV from flight data. The parameters obtained from this class of identification are the weights of the neural network, but these weights cannot be interpreted in a physical sense for subsequent analysis of the UAV system. Another widely used model is the autoregressive moving average with exogenous input (ARMAX) \cite{Babaei2008Sensors} which suffers from the same limitations of neural networks. 

From a data perspective, in some methods, called asymptotic methods, the model parameters are proved to converge in an asymptotic sense \cite{astrm1966_PE} hence theoretically requiring infinite amounts of data. Asymptotic methods are usually formulated in a statistical framework like the maximum likelihood estimators (MLE), the prediction error method (PEM), and various versions of Kalman filters that are used for model parameters' estimation \cite{Hwang2022Sensors}. The infinite amounts of data required by asymptotic methods motivated the interest in system identification techniques with finite sample complexity, like the non-asymptotic methods that require finite amounts of data with guarantees on error bounds. Examples of non-asymptotic methods include the identification of systems based on ordinary least-squares (OLS) for partially observed linear time-invariant (LTI). In \cite{Necmiye2019ACC}, the Markov parameters of the unknown LTI system are estimated using the OLS framework based on a single trajectory of the system. The authors in \cite{Sarkar2022Finite} offered an alternative approach to \cite{Necmiye2019ACC} in which the Hankel matrix of the system is estimated. Both \cite{Necmiye2019ACC} and \cite{Sarkar2022Finite} required the LTI systems to be strictly stable. The work of \cite{Zheng2021multi} extended the methods in \cite{Necmiye2019ACC,Sarkar2022Finite} to unstable systems at the expense of requiring multiple system trajectories. There is a common drawback in all of these asymptotic and non-asymptotic methods in that they do not offer a specific guideline on how data generation (i.e. rolling out trajectories) should be achieved. This results in some practical challenges, like the way data should be generated from unstable or critically stable systems while attaining safety.

Another approach to identification is based on relay testing \cite{astrm1984RFT}, which offers advantages in safety and the use of a standard data generation approach. Due to its simplicity and safety, relay-based testing was adopted in many robotics and automation applications \cite{Boiko2013}. The main drawback of relay-based identification is that they are limited to low-order models, e.g. first-order or second-order models with time delay. These low-order models are usually an approximation of high-order dynamics, which results in issues in the trade-off between robustness and performance that may constrain the desired system performance.  In \cite{Luyben1987}, a first-order system with time delay is identified using a relay feedback test (RFT) while assuming that the static gain of the system is known as apriori. In \cite{Alfaro2021high} the authors developed a method for the identification of high-order models but it is quite restrictive since it is limited to the model with repeated poles and time delay. Another drawback of relay-based identification is the use of the approximate describing function (DF) method which requires the unknown process to exhibit low-pass filtering properties. The use of DF was mitigated, and an exact method based on the locus of the perturbed relay system (LPRS) was suggested for the identification of first-order plus time delay models in \cite{Boiko2008Autotune}, and was extended to higher-order models in \cite{Castellanos2008}. The method in \cite{Castellanos2008}, which is also based on LPRS, requires the solution of a set of equations which would not be feasible in real-time. 

\subsection{Motivation}
System identification of UAV dynamics is a preferred alternative to expensive lab testing. For example, the total drag and the total time delay, which have a significant effect on system dynamics and the required controller, are technically difficult and expensive to obtain through first principles. Moreover, some of the UAV dynamics might change during operation such that re-identification of dynamics would be required using limited onboard sensors, e.g. change of the aerodynamic characteristics or a change of the sensor used for positioning which changes the system delay. Safety and short identification time is a must for uninterrupted missions. Moreover, partial knowledge of some system parameters is easily accessible. For example, propulsion dynamics and the inertia of the system are easy to obtain through bench testing. Knowledge of some parameters a priori would make identification faster and more reliable.

The UAV applications motivate the need for a safe and real-time identification method that accurately estimates the unknown parameters of high-order models given a known model structure and partial knowledge about some parameters of the system.

\subsection{Contribution}
In this work, we propose a method for the identification of high-order LTI time delay systems with known model structure and partial knowledge of some system parameters based on the modified relay feedback test (MRFT). The proposed method inherits the safety and operational convenience of relay-based testing methods, while being applicable in real-time. The real-time capability is achieved by taking advantage of the homogeneous properties of the LPRS which will be used to compute a normalized identification space called the unit frequency manifold (UFM). The UFMs of multiple MRFT tests that will be used onboard the UAV are preloaded in memory for real-time identification. Thus, our contributions can be summarized by the following:
\begin{enumerate}
    \item Proof of homogeneous properties of the LPRS against time and gain scales as well as MRFT parameters scaling.
    \item Construction of an efficient real-time identification algorithm based on the LPRS properties and UFMs.
    \item Experimental validation on multiple multirotor UAVs.
\end{enumerate}

We have applied the proposed identification method to two different UAV platforms. The experimental results show that the identification of each of the dynamic loops (i.e. altitude, roll, or pitch) takes around ten seconds to complete. To validate the identification method accuracy, we intentionally add a known delay in the flight controller software. The identification method correctly estimates the time delay increment. Moreover, time delay and aerodynamic identification results were consistent for loops with the same hardware setup indicating the high precision of the proposed method. A video abstract of the paper is provided in \cite{paper_video}.

\section{Problem Statement}\label{sec_prop_statement}
Let $W_l(s,\mathbf{p})$ be an unknown LTI dynamical system, and let $\mathbf{p}\in \mathbb{R}^N$ be the vector of parameters that characterize \(W_l\), which includes time constants, gains, and delay. Further, let \(\mathbf{\hat{p}}\) be an estimateabout the unknown parameters \(\mathbf{p}\), and let \(E\) and \(I\) represent a set of equality and inequality constraints applied to \(\mathbf{\hat{p}}\), respectively. The set \(E\) represents the pre-known system parameters, e.g. from bench tests or reference manuals. The set \(I\) represents possible pre-known ranges for identification, e.g. range of the unknown mass of a UAV is between 100 g and 10 kg. The constraints \(E\) and \(I\) characterize a subspace \(U \subset \mathbb{R}^N\), such that \(\mathbf{\hat{p}}\in U\).

Since a relay-based test would be used, we further define the relay test as a map (\(\mathbf{p},\mathbf{\zeta})\mapsto (\Omega_t,a_t)=M(\mathbf{p},\mathbf{\zeta})\), where \(\mathbf{\zeta}\) is a vector that characterizes the test parameters, and \(\Omega_t\), \(a_t\) are the frequency and amplitude of the generated limit cycle respectively. The goal of the identification is to find an inverse map \((\Omega_t,a_t,E,I)\mapsto (\mathbf{\hat{p}})=M^{-1}(\Omega_t,a_t,E,I)\) such that \(||\mathbf{p}-\mathbf{\hat{p}}||_n\) is minimized.

In the case of UAV attitude dynamics, a second-order integral plus time delay (SOIPTD) system is to be identified \cite{multirotors}. The nominal model structure is of the form:
\begin{equation}
    \begin{bmatrix}
        \dot{\theta} \\
        \dot{\omega} \\
        \dot{M}  
    \end{bmatrix} = 
    \begin{bmatrix}
        0 & 1 & 0\\
        0 & -\frac{B_x}{J_x}& \frac{1}{J_x}\\
        0 &0 &\frac{1}{T_p} 
    \end{bmatrix}\begin{bmatrix}       
        \theta \\
        \omega \\
        M  
    \end{bmatrix} + \begin{bmatrix}       
        0 \\
        0 \\
        \frac{k_{M,b_x}}{T_p} 
    \end{bmatrix}u(t-\tau_p)
\end{equation}
where $B_x$ is the rotor drag coefficient, $J_x$ is the moment of inertia of the UAV and $k_{M,b_x}$ is the moment gain of the actuator. In the transfer function form, it is written as:
\begin{equation}\label{eq_altitude_attitude_dynamics}
    G(s) = \frac{KT_de^{-\tau s}}{s(T_ps+1)(T_ds+1)}
\end{equation}
where the time constants \(T_p\) and \(T_d = \frac{J_x}{B_x}\) represent the propulsion and aerodynamics time constants respectively, \(\tau\) is the time delay and $K$ is the static gain of the system.

\begin{figure*}[ht]
    \centering
    \includegraphics[width = \textwidth]{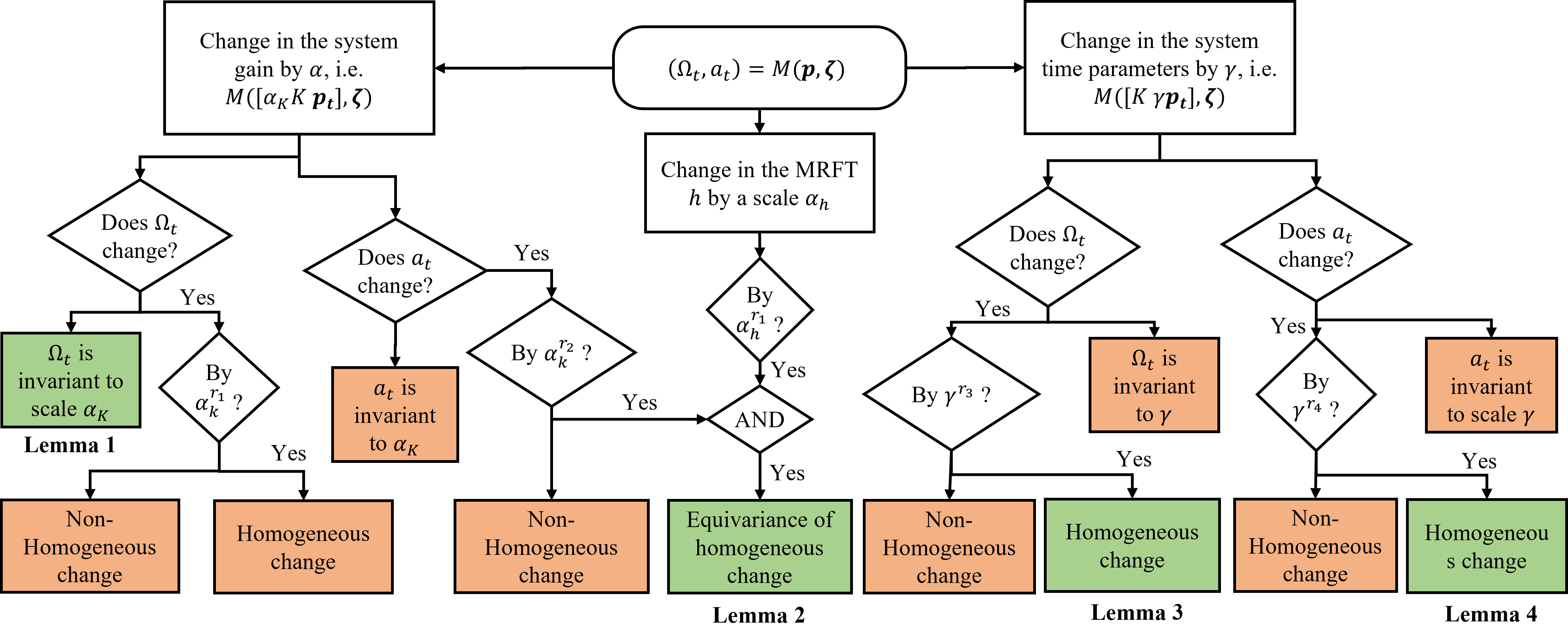}
    \caption{Effect of time and gain scaling of process parameters on MRFT excited oscillations frequency and amplitude. We prove the homogeneity of arbitrary closed-loop linear dynamics with MRFT using the LPRS.}
    \label{fig_conditions_ufm_ugm}
    \vspace{-5mm}
\end{figure*}

\section{Identification on the Normalized Subspace}\label{sec_ufm}
The design of the map \(M\) is essential for the accuracy of estimating \(\mathbf{\hat{p}}\). The definition of \(M\) requires the selection of the relay nonlinearity used in the test algorithm and its tunable parameters \(\mathbf{\zeta}\). For some cases, it is possible to find an analytical solution for \(M^{-1}\) through the use of the Locus of the Perturbed Relay System (LPRS). But for the cases where analytical solutions do not exist, a numerical estimate of \(M^{-1}\) is required. It is possible for the range \(U\) to be an open set, and hence even a numerical estimate of \(M^{-1}\) would not be feasible.

It is yet possible under certain conditions for \(U\) to be mapped to a compact set represented by two special manifolds \(\mathcal{P}\) and \(\mathcal{R}\) which we called the UFM and the \emph{unit gain manifold} (UGM), respectively. Normalizing to unit frequency and unit gain is chosen as a convention. The conditions for the existence of UFM and UGM depend on the properties of the relay used in the test. In this section we introduce MRFT, the test that would be used for data generation, and the conditions for the existance of the UFM and UGM. Finally we discuss the generation of UFM and UGM for the model considered in Eq. \eqref{eq_altitude_attitude_dynamics}.

\subsection{The modified relay feedback test}
The MRFT \cite{Boiko2013} is an algorithm that produces a switching output at a specified phase. The MRFT is given by:
\begin{equation}\label{eq_mrft_algorithm}
u_M(t)=
\left\{
\begin{array}[r]{l l}
h\; &:\; e(t) \geq b_1\; \lor\; (e(t) > -b_2 \;\land\; u_M(t-) = \;\;\, h)\\
-h\; &:\; e(t) \leq -b_2 \;\lor\; (e(t) < b_1 \;\land\; u_M(t-) = -h)
\end{array}
\right.
\end{equation}
where \(b_1 = -\beta e_{min}\) and \(b_2 = \beta e_{max}\), and \(u_M(-t)\), \(e_{max}\), and \(e_{min}\) are the previous command, maximum error, and minimum error, respectively. MRFT has tunable parameters \(\bm{\zeta}=[\beta\;h]^T\) which needs to be designed. The describing function (DF) of MRFT is presented in \cite{Boiko2013} as:
\begin{equation}\label{eq_mrft_df}
N_M(a)=\frac{4h}{\pi a}(\sqrt{1-\beta^{2}}-j\beta)
\end{equation}

Using the DF, it could be shown that the MRFT achieves oscillations at a specified phase angle by satisfying the HB equation \cite{Boiko2013}:
\begin{equation}\label{eq_hb}
W_l(j{\Omega}_t)=\frac{-1}{N_M({a}_t)}
\end{equation}
with the RHS given by:
\begin{equation}
    \frac{-1}{N_M({a}_t)}=\frac{-\pi {a}_t}{4h}\left(\sqrt{1-\beta^2} +j\beta\right) 
\end{equation}
and the corresponding magnitude and phase of the RHS of Eq. \eqref{eq_hb} given by:
\begin{equation}\label{eq_mrft_hb_mag_phase}
    \begin{aligned}
        |\frac{-1}{N_M({a}_t)}|=\frac{\pi {a}_t}{4h}\\
        \arg \frac{-1}{N_M({a}_t)}= -\pi+\arcsin{\beta}
    \end{aligned}
\end{equation}

Note that the existence of a stable limit cycle for the UAV model when using MRFT was proven in \cite{ayyad2021multirotors}, and is assumed hereafter. The predicted limit cycle frequency \({\Omega}_t\) and amplitude \({a}_t\) are exact since the LPRS is adopted in the present paper.

\vspace{-3mm}
\subsection{Proof of homogeneity properties using the LPRS}\label{proof_lemmas}

The conditions for the generation of UFM and UGM is governed by the four lemmas summarized in Fig. \ref{fig_conditions_ufm_ugm}. Lemmas 1 and 2 would allow the system parameters to be split \(\mathbf{p}=[K\;\mathbf{p_t}]^T\) where \(\mathbf{p_t}\) are the time-dependent parameters of the system. Lemmas 3 and 4 relate the generated test frequency with the unknown system time parameters. To prove these lemmas we use the LPRS. 

The LPRS is a method that was introduced in \cite{BOIKO2005677} which can provide an exact solution of the periodic motion in a relay feedback system. A variation of LPRS, denoted as the \(\Phi(\omega)\) function \cite{boiko2007analysis}, can be used to provide the exact solution for oscillations when the system is excited by MRFT. The complex function \(\Phi(\omega)\) was designed to be compatible with the HB formulation so that the MRFT DF can be used to predict limit cycles. The LPRS based approach for analysis of periodic motion is achieved by replacing the system frequency response \(W_l(j\Omega_t)\) in the HB Eq. \eqref{eq_hb} by \(\Phi\) to obtain:
\begin{equation}\label{eq_exact_hb}
    \Phi(\Omega_t)=\frac{-1}{N_M(a_t)}
\end{equation}
though the DF of MRFT is used, the predicted limit cycle frequency \(\Omega_t\) and amplitude \(a_t\) are exact since \(\Phi(\omega)\) accounted for all harmonics in the relay switching. The Lemmas 1-4 will be proven using \(\Phi(\omega)\). The complex function \(\Phi(\omega)\) is given by:
\begin{equation}\label{eq_phi_mrft}
    \Phi(\omega)=-\sqrt{[a_y(\omega)]^2-y^2(\frac{\pi}{\omega},\omega)}+jy(\frac{\pi}{\omega},\omega)
\end{equation}
where \(y(\frac{\pi}{\omega},\omega)\) is the value of the system periodic output at the instant of switch of MRFT from \(-h\) to \(h\).  The amplitude of the process output \(a_y(\omega)\) is given by:
\begin{equation}\label{eq_amplitude_mrft_exact}
    a_y(\omega)=\max_{t\in[0,T]}|y(t,\omega)|
\end{equation}

Note that \(y(\frac{\pi}{\omega},\omega)=\beta a_y(\omega)\). The periodic output signal \(y(t,\omega)\) may be found using the Fourier series:
\begin{equation}\label{eq_fourier_series_mrft}
\begin{aligned}
y(t,\omega)=\frac{4h}{\pi}\sum_{k=1}^\infty & \frac{(-1)^{k+1}}{2k-1}\sin[(2k-1)\omega t+\\
&\varphi_l((2k-1)\omega)]A_l((2k-1)\omega)
\end{aligned}
\end{equation}
where \(\varphi_l(\omega)=\arg W_l(j\omega)\), \(A_l(\omega)=|W_l(j\omega)|\). Assuming a nominal system \(W_{l0}\) given by:
\begin{equation}
    W_{l0}(j\omega_0)=K_0\frac{\prod_i(jT_{N_i}\omega_0+1)e^{-j\tau \omega_0}}{(j\omega_0)^{n_i}\prod_j(jT_{D_j}\omega_0+1)}
\end{equation}
with corresponding function \(\Phi_0(\omega_0)\), amplitude of oscillation \(a_{y,0}(\omega_0)\), and output \(y_0(t,\omega_0)\). The magnitude \(|W_{l0}(j\omega_0)|\) is then given by:
\begin{equation*}
    |W_{l0}(j\omega_0)|=K_0\frac{\prod_i\sqrt{(T_{N_i}\omega_0)^2+1}}{(\omega_0)^{n_i}\prod_j\sqrt{(T_{D_j}\omega_0)^2+1}}
\end{equation*}
and the phase \(\arg W_{l0}(jk\omega_0)\) is given by:
\begin{align*}    
    \arg W_{l0}(j\omega_0)=\sum_i\arctan{T_{N_i}  \omega_0}-\sum_i\arctan{T_{D_i}  \omega_0}\\-n_i\frac{\pi}{2}-\tau \omega_0
\end{align*}

\paragraph*{\textbf{Lemma 1}}  The frequency $\Omega_t$ of the test oscillations under the MRFT is invariant to the change of the static gain of the system

\paragraph*{Proof of Lemma 1}

Let $\Omega_{t0}$ be the frequency of the test oscillations corresponding to the process \(W_{l0}\). Further let \(W_{l\alpha}(j\omega_\alpha)\) be a system that is related to the nominal system \(W_{l0}\) by a scaled static gain \(K_\alpha=\alpha_K K_0\). Since  \(\arg W_{l\alpha}(j\omega_\alpha)\) is independent of \(K_\alpha\), we get \(\arg W_{l\alpha}(j\omega_\alpha)=\arg W_{l0}(j\omega_0)\). This leads to the phase balance of Eq. \eqref{eq_exact_hb}, which leads to the invariance of test frequency \(\Omega_{t\alpha}=\Omega_{t0}\). 

Moreover, changing the MRFT amplitude \(h\) in Eq. \eqref{eq_mrft_hb_mag_phase} does not change the limit cycle phase, and hence the phase balance in the Eq. \eqref{eq_exact_hb} remains the same, leading to invariance of the test frequency \(\Omega_{t0}\). Hence, Lemma 1 is proved. 

\paragraph*{\textbf{Lemma 2}}  The amplitude $a_t$ of the test oscillations under the MRFT is a homogeneous function of the system static gain and MRFT amplitude

\paragraph*{Proof of Lemma 2}
Given the magnitude relationship \(|W_{l\alpha}(j\omega_0)|=\alpha|W_{l0}(j\omega_0)|\), we obtain \(y_\alpha(t,\omega_0)=\alpha y_0(t,\omega_0)\) for Eq. \eqref{eq_fourier_series_mrft}, and \(a_{y,\alpha}=\alpha a_{y,0}\) for Eq. \eqref{eq_amplitude_mrft_exact}. Then, using Eq. \eqref{eq_phi_mrft}, the LPRS \(\Phi_\alpha(\omega_0)\) is related to \(\Phi_0(\omega_0)\) by:
\begin{multline}
    \Phi_\alpha(\omega_0)=-\sqrt{\alpha^2[a_{y,0}(\omega_0)]^2-\alpha^2y_0^2\left(\frac{\pi}{\omega},\omega_0\right)}+\alpha jy_0(\frac{\pi}{\omega_0},\omega_0)\\=
    \alpha(-\sqrt{[a_{y,0}(\omega_0)]^2-y_0^2(\frac{\pi}{\omega_0},\omega_0)}+jy_0(\frac{\pi}{\omega_0},\omega_0))=\alpha\Phi_0(\omega_0)
    \label{eq_gain_scale_mag}
\end{multline}
which is homogeneous with degree one. 

Note that scaling the relay amplitude in the RHS of Eq. \eqref{eq_exact_hb} as \(h_{\alpha_h}=\alpha_h h_0\) would simply result in \(a_{t,\alpha_h}=\alpha_h a_{t,0}\) which is also homogeneous with degree one. Since both scaling the system gain and the MRFT amplitude resulted in the same homogeneity degree, there is an equivariance between these scales, which proves Lemma 2.

\paragraph*{\textbf{Lemma 3}}  The frequency $\Omega_t$ of the test oscillations under the MRFT is a homogeneous function of the system time parameters

\paragraph*{Proof of Lemma 3} Consider the system \(W_{l\gamma}(j\omega_\gamma)\) related to the nominal system \(W_{l0}(j\omega_0)\) by the scale of the system time parameters \(\bm{p}_\gamma=\gamma\bm{p}_0\). Given the phase equation of \(W_{l\gamma}(j\omega_\gamma)\):
\begin{align*}
    \arg W_{l\gamma}(j\omega_\gamma)=\sum_i\arctan{\gamma T_{N_i} \omega_\gamma}-\sum_i\arctan{\gamma T_{D_i} \omega_\gamma}\\-n_i\frac{\pi}{2}-\gamma\tau \omega_\gamma
\end{align*}
and assuming the new frequency \(\omega_\gamma\) is related to the nominal system frequency by:
\begin{equation}\label{eq_time_scale_effect}
    \omega_\gamma=\frac{\omega_0}{\gamma}
\end{equation}
we obtain \(\arg W_{l\alpha}(j\frac{\omega_0}{\gamma})=\arg W_{l0}(j\omega_0)\), which leads to the homogeneous relation of the test frequency \(\Omega_\gamma=\frac{\Omega_0}{\gamma}\), and therefore, proves Lemma 3.

\paragraph*{\textbf{Lemma 4}}  The amplitude $a_t$ of the test oscillations under the MRFT is a homogeneous function of the system time parameters

\paragraph*{Proof of Lemma 4} The magnitude of the system \(W_{l\gamma}(j\omega_\gamma)\) is related to the nominal system by:
\begin{align}
    |W_{l\gamma}(j\omega_\gamma)|&=\gamma^{n_i}K_0\frac{\prod_i\sqrt{(\gamma T_{N_i}\omega_\gamma)^2+1}}{(\omega_\gamma)^{n_i}\prod_j\sqrt{(\gamma T_{D_j}\omega_\gamma)^2+1}}
\end{align}
and using the result from Lemma 3 proof:
\begin{align}
    |W_{l\gamma}(j\frac{\omega_0}{\gamma})|&=\gamma^{n_i}K_0\frac{\prod_i\sqrt{(T_{N_i}\omega_0)^2+1}}{(\omega_0)^{n_i}\prod_j\sqrt{(T_{D_j}\omega_0)^2+1}}\\
    &=\gamma^{n_i}|W_{l0}(j\omega_0)|
\end{align}
which leads to \(y_\gamma(\gamma t,\frac{\omega_0}{\gamma})=\gamma^{n_i}y_0(t,\omega_0)\) and the LPRS of the nominal and time-scaled systems are related by (we use the results from Eq. \eqref{eq_gain_scale_mag}):
\begin{equation}
    \Phi_\gamma(\frac{\omega_0}{\gamma})=\gamma^{n_i}\Phi_0({\omega_0})
\end{equation}

Hence, the amplitude of the test is homogeneous with respect to time scaling with degree \(n_i\), which proves Lemma 4.

\vspace{-3mm}
\subsection{Unit frequency manifold}\label{UFM}
Achieving parameter identification in the unbounded parameter space in real-time is not feasible. To overcome this, a bounded normalized parameter space is obtained by leveraging the homogeneity properties of MRFT. 
If the time parameters of a nominal process are scaled by a factor $\gamma$, the frequency of oscillation also scales by a factor of $\frac{1}{\gamma}$. From this result, we may argue that for every process in the parameter space that produces oscillations of frequency $\Omega_0$, we can find a corresponding process that produces oscillations at an arbitrary frequency $\hat{\Omega}$ by choosing $\gamma = \frac{\Omega_0}{\hat{\Omega}}$. Using this, we find a set of process parameters that produce the same frequency; the identification will be done on this set. The set of processes that produce any other frequency of oscillations can be obtained by simple scaling. By convention, we choose all the processes on the set to have oscillations of 1Hz frequency, and therefore, we call this set the \textit{unit frequency manifold} (UFM). Similarly, by Lemma 1, due to the invariance of the frequency to the static gain of the system, the processes in the UFM may have different static gains. By convention, we define the static gain of all the processes in the UFM to be unity. Therefore, the set of amplitudes of the processes in the UFM is called \textit{unit gain manifold} (UGM). 

Introducing the idea of UFM and UGM offers us many practical advantages. Firstly, for a class of LTI systems, the UFM and UGM can be computed beforehand, which allows the identification to be performed in real-time. This was not feasible before in the LPRS-based identification. Secondly, the parameter space is compacted to a surface which would decrease the computing power and memory required for identification. 

\begin{figure}
     \centering
     \begin{subfigure}[b]{0.45\textwidth}
         \centering
         \includegraphics[width=\textwidth]{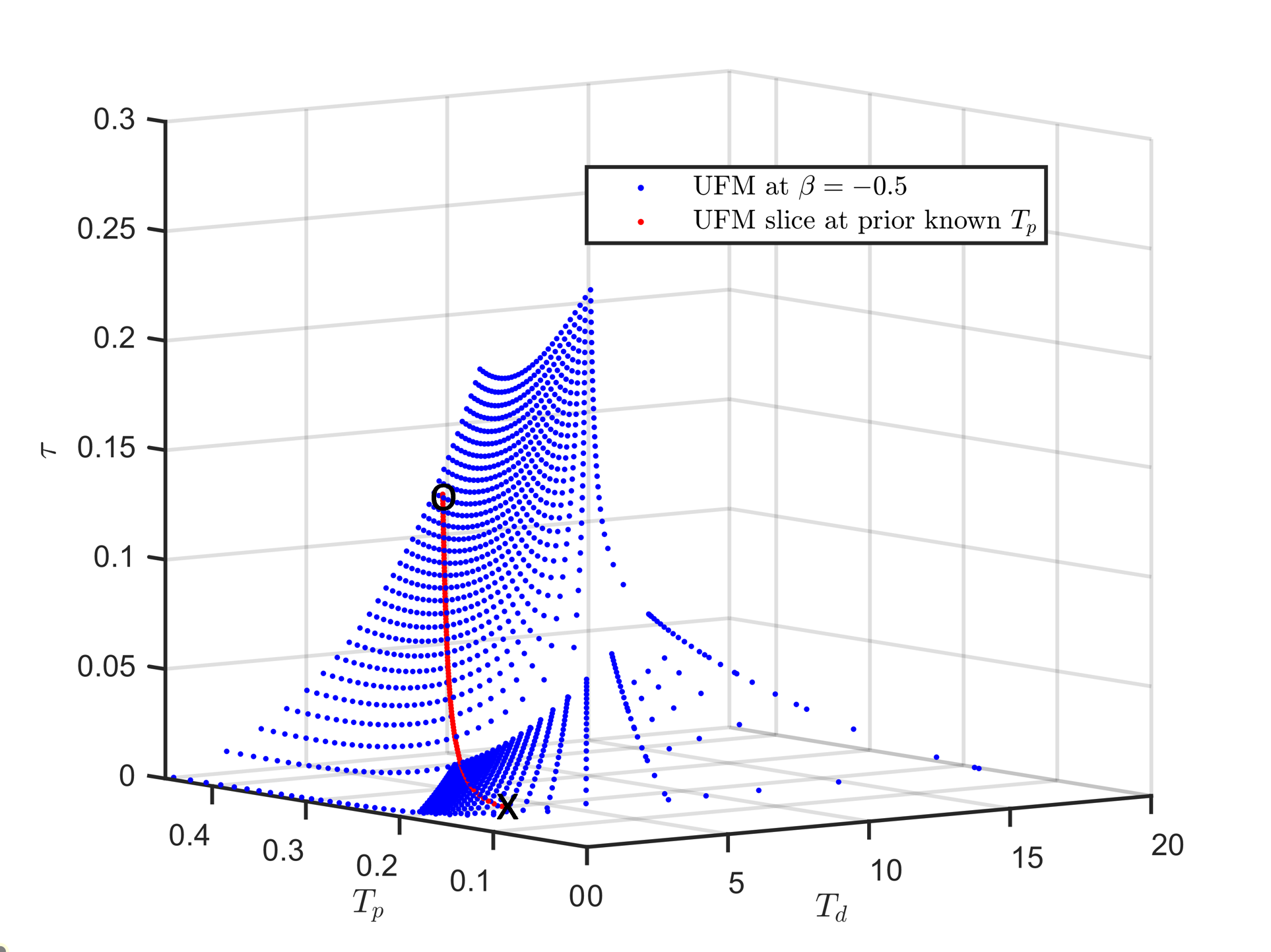}
         \caption{Unit frequency manifold for $\beta = -0.5$}
         \label{fig:UFM}
     \end{subfigure}
     \hfill
     \begin{subfigure}[b]{0.45\textwidth}
         \centering
         \includegraphics[width=\textwidth]{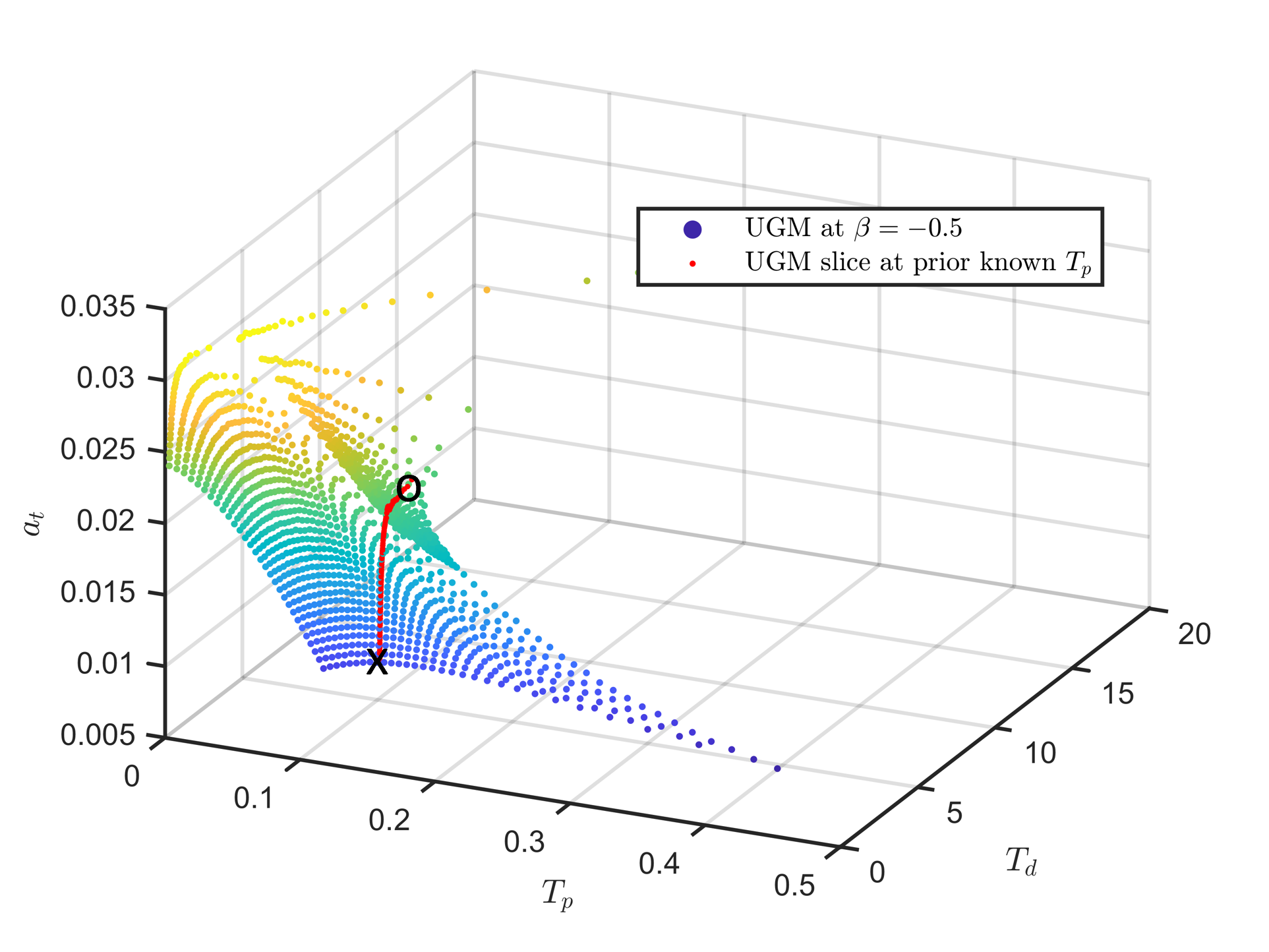}
         \caption{Unit gain manifold for $\beta = -0.5$}
         \label{fig:UGM}
     \end{subfigure}
     \caption{A slice of the UFM which corresponds to the known $T_p$ is shown in red in Fig. \ref{fig:UFM}. The amplitude of oscillations for this set of processes is shown in 
    Fig. \ref{fig:UGM}.}
     \hfill
     \vspace{-7mm}
\end{figure}

Consider the SOIPTD model of the UAV dynamics as described in Eq. \eqref{eq_altitude_attitude_dynamics}.
The UFM and the corresponding UGM of this model for a particular beta were computed and illustrated in Fig. \ref{fig:UFM} and Fig. \ref{fig:UGM}.
It was found through simulation tests that using UGM for identification gives inaccurate results in the presence of noise. Therefore, the identification is done solely based on the frequency of oscillations. We require two oscillations from MRFT to identify the parameters of a UAV.
The simulation results that show the inaccuracies from using the UGM are further discussed in Section \ref{sim_results}.
\section{UAV Model}\label{sec_uav_case}
\subsection{Reference Frames and Conventions}
We define an inertial frame \(\mathcal{F}_I\) having basis \( \bm{[i_x, i_y, i_z]} \) with \(\bm{i_z}\) antiparallel to the gravity vector, and a body-fixed reference frame \(\mathcal{F}_B\) centered at the center of gravity of the UAV with rotation matrix \(  {}^I_B\bm{R} = \bm{[b_x, b_y, b_z]} \in \text{SO(3)} \), which gives the transformation from \(\mathcal{F}_B\) to \(\mathcal{F}_I\), where \( \bm{b_z} \) is parallel to the thrust vector. We also define the horizon frame $\mathcal{F}_H$ with its origin coincident with the origin of $\mathcal{F}_I$, its basis \(\bm{h_z}\) being coincident with \(\bm{i_z}\), and it is yaw aligned with $\mathcal{F}_B$. A vector can be expressed in a particular reference frame, e.g. \({}^I\bm{p}\) is the position vector expressed in the inertial frame. The components of a vector are referred to with the subscripts as in \({}^I\bm{p}=[{}^Ip_x {}^Ip_y {}^Ip_z]^T\). For compatibility of notation with vector quantities we use \(K_x\) to represent the element \(K_{11}\) in a diagonal matrix, and so on.

\subsection{Nonlinear Time Delay UAV Model}
\label{sec_time_delay_model}
We define the motor commands as follows:
\begin{equation}
    \begin{bmatrix}
    \bm{u_\eta} \\
    u_T
    \end{bmatrix} = \bar{G}\bm{u_i}
\end{equation}
where \(\bm{u_\eta}=[u_{b_x} u_{b_y} u_{b_z}]^T\) represents torque commands around \(\mathcal{F}_B\) bases, \(u_T\) is the thrust command, and \(\bm{u_i}\in[0,1]\) is the dimensionless individual motor command with \(i\in \{1,...,\mu_n\}\) where \(\mu_n\) represents the number of propellers used. \(\bar{G}\in\mathbb{R}^{4\times \mu_n}\) provides a static map independent of UAV dynamics, with \(\text{rank}(\bar{G})=4\) and its Moore–Penrose inverse \(\bar{G}^{+}\) is defined and unique. The individual propulsion system thrust and moment dynamics are given by:
\begin{equation}
\label{eq_act_dynamics_time}
\begin{aligned}
    F_i(t) &= k_F u_i(t-\tau_p)-T_p \dot{F}_i(t) \\
    M_i(t) &= k_M u_i(t-\tau_p)-T_p \dot{M}_i(t)
\end{aligned}
\end{equation}
where \(k_F\), \(k_M\), \(\tau_p\), and \(T_p\) are the thrust gain, moment gain, propulsion system time delay, and propulsion system time constant respectively. Note that we assume that all propulsion units are matched, i.e. the parameters \(k_F\), \(k_M\), \(\tau_p\), and \(T_p\) are the same for all rotors. Also, it is assumed that the thrust and moments applied to the rigid body are defined by the relation:
\begin{equation}
    \begin{bmatrix}
    \bm{M} \\
    F
    \end{bmatrix} = G_F\bm{F_p}+G_M\bm{M_p}
\end{equation}
where \(\bm{F_p}=[F_1 F_2 ... F_{\mu_n}]^T\) and \(\bm{M_p}=[M_1 M_2 ... M_{\mu_n}]^T\). \(G_F\) and \(G_M\) are static maps which may contain UAV dynamic parameters. The UAV body dynamics are then given by:
\begin{equation}
\label{eq_uav_dynamics}
    \begin{aligned}
        {}^I\dot{\bm{p}}&={}^I\bm{v}\\
        {}^I\dot{\bm{v}}&=-g\bm{i_z}+\frac{F}{m}\bm{b_z}-{}_B^I{R}D{}_I^B{R}{}^I\bm{v}\\
        \dot{R}&={}_B^IR{}^B\bm{\omega}\\
        \bm{\dot{\omega}}&=J^{-1}(\bm{M}-\bm{\omega}\times J\bm{\omega}-\bm{M_g}-A{}_I^B{R}{}^I{\bm{v}}-B\bm{\omega})
    \end{aligned}
\end{equation}
where the diagonal matrices \(D\), \(J\), \(A\) and \(B\) represents profile drag and inflow motion drag due to translational motion, moment of inertia, drag due to blade flapping, and rotational drag due to body profile and inflow motion, respectively. The vector \(\bm{M_g}\) represents gyroscopic moments due to the interaction between rotating propellers and rotating UAV body.

\vspace{-2mm}
\subsection{Dynamics decoupling}
Analysis of decoupled dynamics can be achieved by projecting the 3D space into a 2D space. Specifically for decoupling, we assume that \(\mathcal{F}_I:=\mathcal{F}_H\) and, without loss of generality, project on the plane defined by \(\bm{i_y}\times \bm{i_z}\). The rotation around \(\bm{b_x}\) is indicated by the angle \(\theta\). We use near-hover linearization assumptions, i.e. we linearize about zero pitch and roll angles, and we use linear drag models. The rotational dynamics become:
\begin{equation}\label{eq_attitude_linearized}
    \begin{aligned}
        \dot{\theta}&=\omega_x\\
        \dot{\omega}_x&=\frac{1}{J_x}(M_x-A_x{}^Iv_y-B_x\omega_x)\\
        \dot{M}_x&=\frac{-M_x+k_{M,b_x}u_{b_x}(t-\tau_p)}{T_p}
    \end{aligned}
\end{equation}

We assume that the contribution of the term \(A_x{}^Iv_y\) is small since the lateral velocity during identification near-zero, and hence we neglect it. Then the angular dynamics take the structure of Eq. \eqref{eq_altitude_attitude_dynamics} with the model parameters given by:
\begin{equation}\label{eq_parameters_for_attitude}
    \begin{aligned}
        T_d&=\frac{J_x}{B_x}\\
        \tau&=\tau_p+\tau_{imu}
    \end{aligned}
\end{equation}
where $\tau_{imu}$ is the time delay that arises from the IMU measurement and the real-time processing.

The dynamics of altitude loop on the plane \(\bm{i_y}\times \bm{i_z}\) is given by:
\begin{equation}
    \begin{aligned}
    {}^I\dot{p}_z=&{}^Iv_z\\
    {}^I\dot{v}_z=&\cos\theta a_F - (d_y\sin\theta\cos\theta+d_z\sin\theta \cos\theta){}^Iv_y \\
    &- (d_y\sin^2\theta+d_z\cos^2\theta){}^Iv_z\\
    \dot{a}_F=&\frac{\mu_n\frac{k_F}{m}u_T(t-\tau_p)}{T_p}-\frac{a_F}{T_p}
    \end{aligned}
\end{equation}
where $a_F = F/m$.
Given the fact that we perform identification on altitude with \(\theta\approx0\) the altitude dynamics become:
\begin{equation}
    \begin{aligned}
    {}^I\dot{p}_z=&{}^Iv_z\\
    {}^I\dot{v}_z=& a_F -d_z{}^Iv_z\\
    \dot{a}_F=&\frac{\mu_n\frac{k_F}{m}u_T(t-\tau_p)}{T_p}-\frac{a_F}{T_p}
    \end{aligned}
\end{equation}
and thus, similar to the angular dynamics, the altitude dynamics take the form of Eq. \eqref{eq_altitude_attitude_dynamics} with the model parameters given by:
\begin{equation}\label{eq_parameters_for_altitude}
    \begin{aligned}
        T_d&=\frac{1}{D_z}\\
        \tau&=\tau_p+\tau_{pos}
    \end{aligned}
\end{equation}
Where $\tau_{pos}$ includes the time delay that comes from the position sensor as well as the onboard flight computer used.

\section{Results}\label{sec_results}
\subsection{Simulation results}\label{sim_results}
The proposed identification method is first validated in simulation since we have access to the ground truth process parameters. Consider the UAV attitude dynamics given by:
\begin{equation}\label{simulated_example}
    \begin{bmatrix}
        \dot{\theta} \\
        \dot{\omega} \\
        \dot{M}  
    \end{bmatrix} = 
    \begin{bmatrix}
        0 & 1 & 0\\
        0 & -1.42& 1.42\\
        0 &0 & 10 
    \end{bmatrix}\begin{bmatrix}       
        \theta \\
        \omega \\
        M  
    \end{bmatrix} + \begin{bmatrix}       
        0 \\
        0 \\
        1.4 
    \end{bmatrix}u(t-0.06)
\end{equation}
To assess the sensitivity of the identification method, white Gaussian noise was added to the frequency and amplitude obtained from simulation to study the effect of noisy measurements on the identified UAV parameters. First, the system in Eq. \eqref{simulated_example} was excited by MRFT with $\beta=-0.7$, which resulted in oscillations with frequency and amplitude of $1.022$ Hz and $0.05$, respectively. With the addition of white noise with a standard deviation of 3\% of the resultant amplitude and frequency, the mean $T_d$ was found to be 0.39 s with a standard deviation of 0.089 s, a 44\% error from the true value. Similarly, the mean $\tau$ was found to be 0.12 s with a standard deviation of 0.0133 s, a 100\% error from the true value. This shows that the UGM exhibits an asymmetric nonlinear effect leading to high biases. Therefore, the amplitude of the oscillations can hardly be used for accurate identification.

This necessitates the use of the frequencies of two MRFT tests in identification to avoid the use of the amplitude of the oscillations. We excited the system in Eq. \eqref{simulated_example} with two MRFTs of $\beta = -0.4$ and $\beta = -0.7$ which produced oscillations with frequencies of $0.708$ Hz and $1.022$ Hz respectively. A similar sensitivity analysis was conducted for the identification based on the two frequencies where a white noise with standard deviation of 3\% was added to the two resultant frequencies, which resulted in the identification with mean $T_d = 0.6404$ s and standard deviation of $0.1956$ s, and for the time delay, we have mean $\tau = 0.0679$ s with standard deviation of $0.0168$ s. The error percentage in this case was 8.5\% and 6.73\% for $T_d$ and $\tau$ respectively, which is significantly lower than the case when the amplitude of the oscillations was considered in the identification. Therefore, we only depend on frequency measurements for experimental identification. The computations required for the identification using the two frequencies case was less than $0.1$ seconds on common modern processors, leading to real-time applicability.



\subsection{Experimental setup}
We validate the proposed identification methodology experimentally using two different UAV platforms. Furthermore, we have developed a test rig for the estimation of UAV propulsion system dynamics which will be used as prior knowledge for MRFT-based identification.

\subsubsection{UAV platform}
\begin{figure}
    \centering
    \includegraphics[width = 0.7\columnwidth]{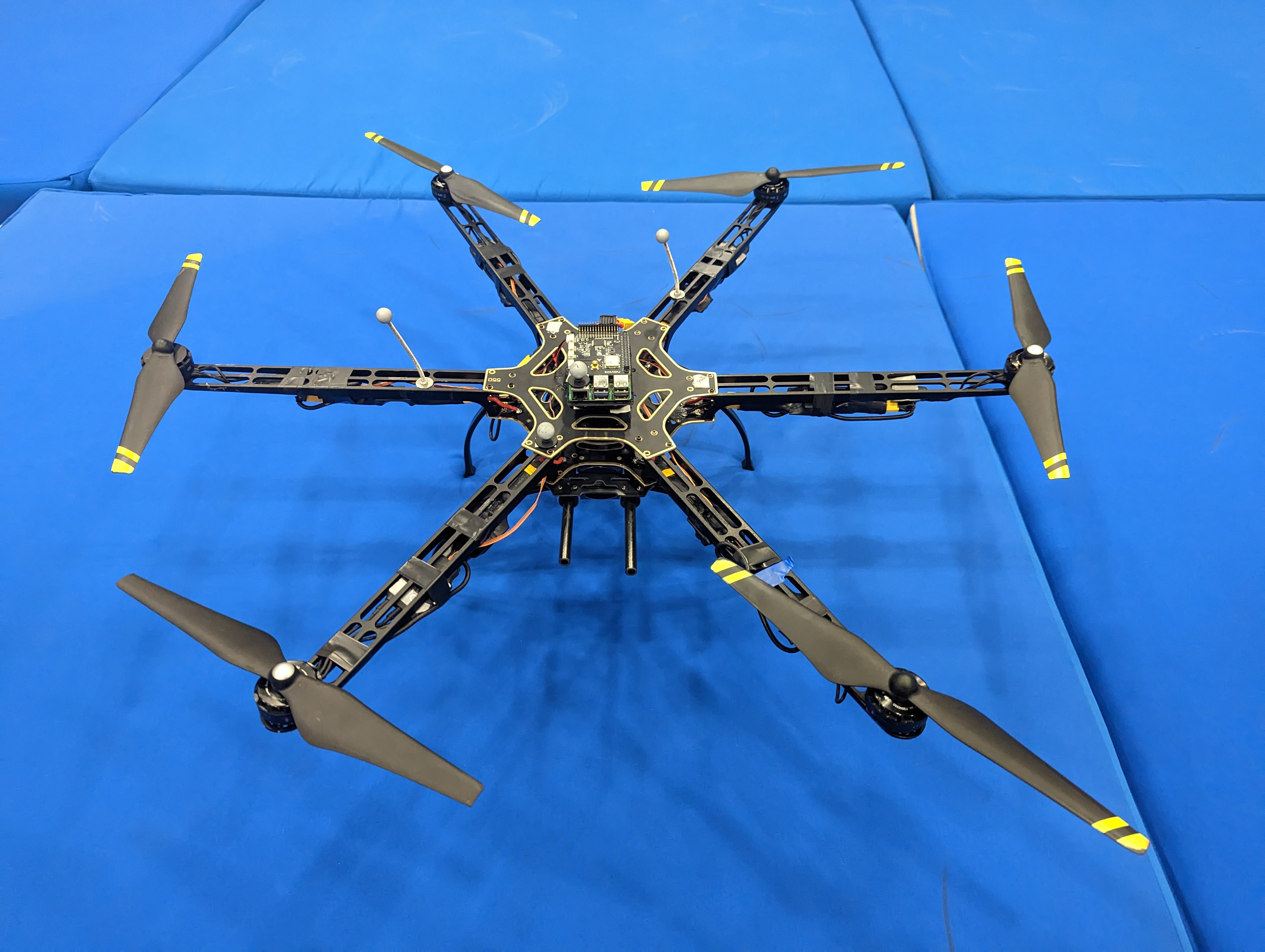}
    \caption{The experimental hexarotor UAV Design II.}
    \label{fig:flight_env}
\end{figure}

The UAV platform used for the experiments is a dji F550 hexarotor that uses a Navio2 flight controller with Raspberry Pi 3B+. Two UAV designs equipped with \textit{TDK-Lambda i7A} DC-DC voltage regulators on their electric power train are used for experimentation. The voltage regulators fix the static gain of the system; therefore, the battery voltage drop no longer affects propulsion system gain. In \emph{UAV Design I} dji E305 propulsion system is used, and in \emph{UAV Design II} dji E600 propulsion system is used. The position and yaw states of the UAVs are measured using OptiTrack motion capture system with sampling at 200Hz, and the roll and pitch measurements of the UAV are obtained from the onboard IMU with sampling at 200Hz. The communication between the ground station and the flight controller is done over a WiFi network, with ROS being used as a middleware. The UAV Design II used in the experiments is shown  Fig. \ref{fig:flight_env}.

\subsubsection{Test rig for the estimation of propulsion system dynamics}\label{test_bench}
The altitude dynamics of the UAV system consist of two time constants and a time delay. The time constant corresponding to aerodynamic drag and the time delay are expensive to test offline. Therefore, the time constant of the propulsion system is identified offline on a test rig, and the other two parameters will be identified online based on the MRFT oscillations. 

To test propulsion dynamics, we prepared a fulcrum balance that is free to rotate around its center. We placed the propulsion system on one side of the balance and \textit{Kistler 9272} force sensor with a high sampling rate of 1KS/s on the other side of the balance. A step command is then used as an input to the propulsion system and the resultant force is then observed by the force sensor.

\begin{figure}
    \centering
    \includegraphics[width = \columnwidth]{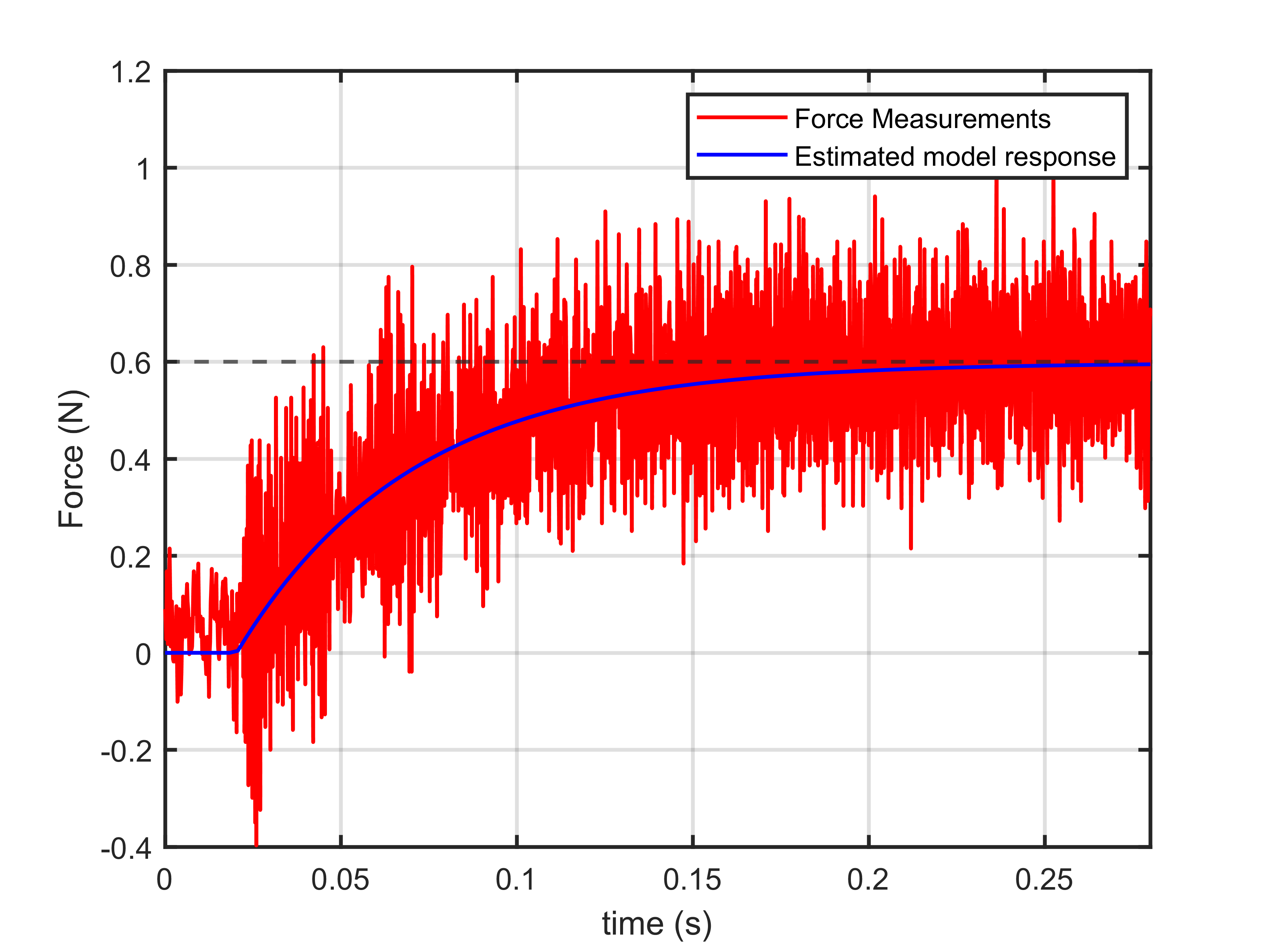}
    \caption{The identified propulsion dynamics are compared to the step test of the actual propulsion system}
    \label{fig:act_identified}
\end{figure}

For the two propulsion systems used in the experimentation, a least mean squares method was used to fit the parameters of the propulsion system to the measurement data. The parameters thus obtained for the two propulsion systems are tabulated in Table \ref{tab:act_params} and the step response of the identified propulsion dynamics model is compared with the measured propulsion force in Fig. \ref{fig:act_identified}.

\begin{table}
    \centering
    \begin{tabular}{|c|c|c|}
    \hline
         Actuator system & $T_P$ & $\tau$ \\\hline
         Dji E305 (used in UAV Design I) & 0.0422 & 0.017 \\
         Dji E600 (used in UAV Design II) & 0.0499 & 0.0203\\\hline
    \end{tabular}
    \caption{The parameters identified for different propulsion systems}
    \label{tab:act_params}
    \vspace{-5mm}
\end{table}


\subsection{Real-time identification results}

The identification is performed on each control loop separately. Altitude is tested with MRFT of $\beta$ values of $-0.4$ and $-0.7$, and attitude is tested with beta parameters of $\beta =-0.5$ and $-0.8$. The MRFT parameters are chosen to be spread out as much as possible within the constraints of the amplitude and frequency permissible in the experimental setup. For the attitude dynamics, higher values of $\beta$ are used because lower values of $\beta$ produced oscillations with amplitudes that are dangerously high.

\begin{figure}
     \centering
     \vspace{-10mm}
     \begin{subfigure}[b]{0.43\textwidth}
         \centering
         \includegraphics[width=\textwidth]{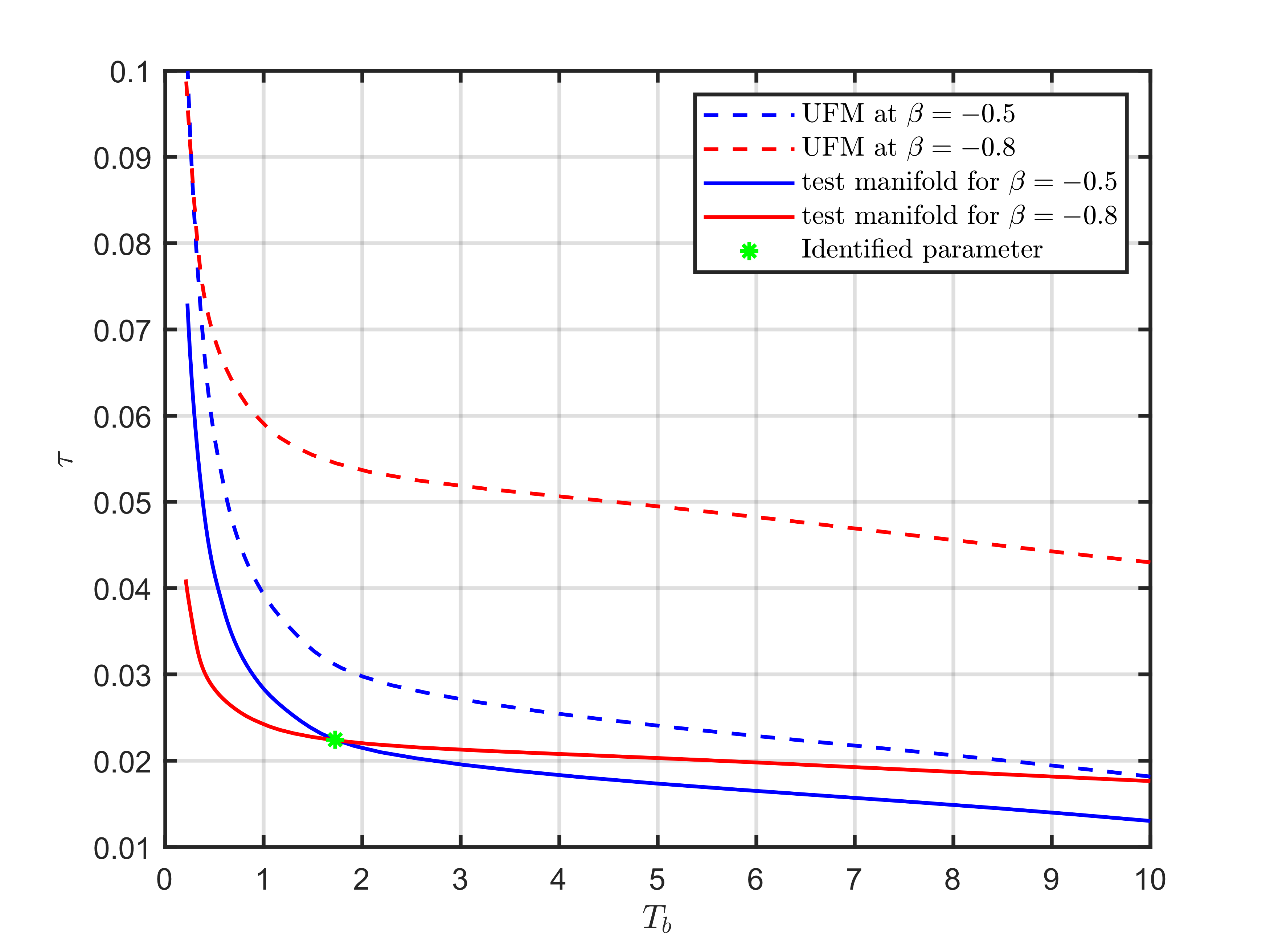}
         \caption{Identification of the pitch parameters based on the intersection of the two manifolds}
         \label{fig:pitch_identification}
     \end{subfigure}
     \hfill
     \begin{subfigure}[b]{0.43\textwidth}
         \centering
            \includegraphics[width = \textwidth]{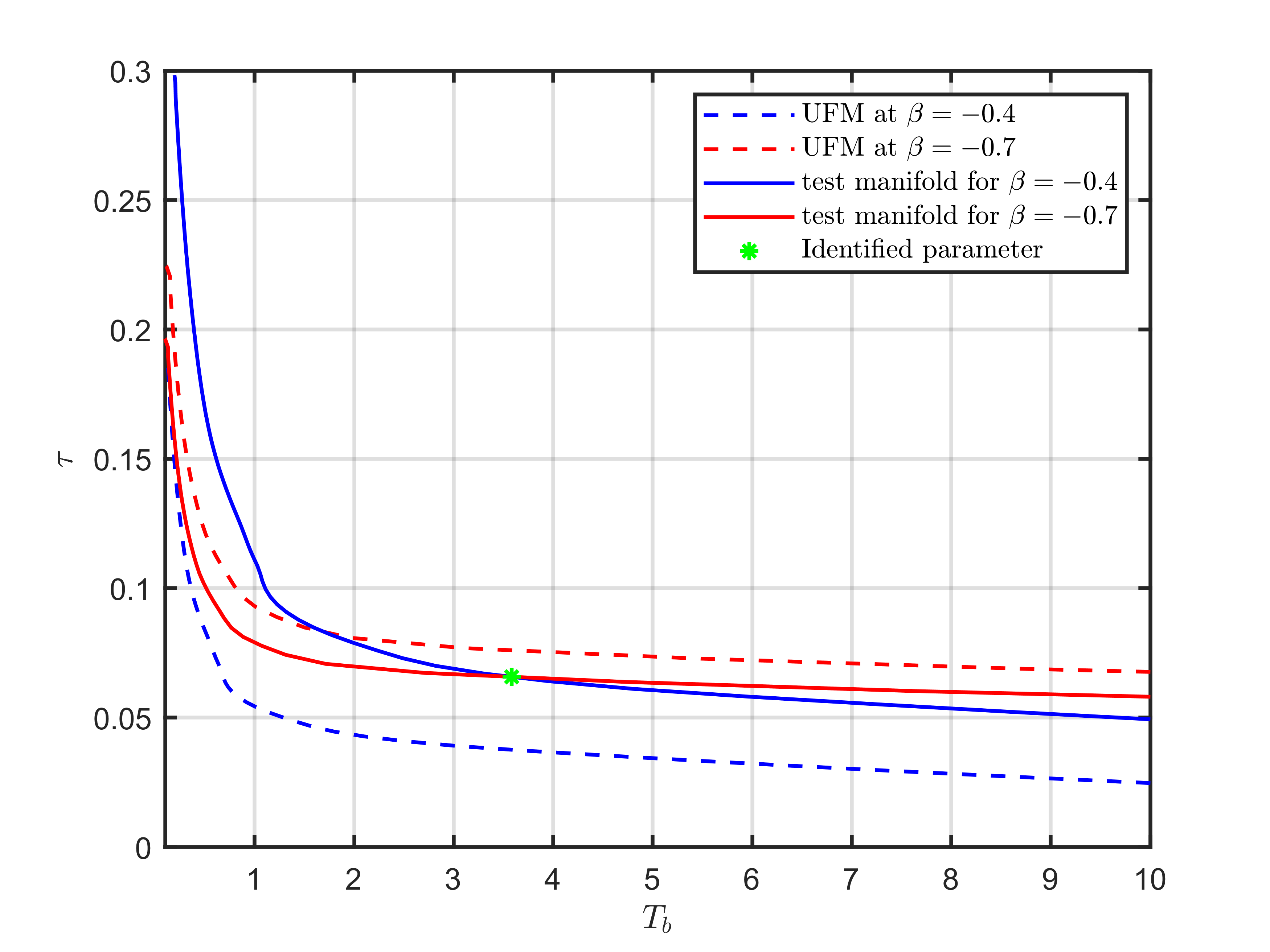}
            \vspace{-8mm}
        \caption{The parameters identified for the altitude dynamics of the system with additional 20 ms added are given by the point of intersection of the two manifolds.}
        \label{fig:alt_identified}
     \end{subfigure}
     \caption{The parameters for the altitude dynamics as well as the attitude dynamics is obtained from the intersection point of the two manifolds}
     \label{fig_pitch_alt_ident}
\end{figure}

The frequencies of the two oscillations obtained for altitude dynamics are 0.63 Hz and 1.1 Hz for $\beta=-0.4$ and $\beta=-0.7$, respectively. The known $T_p$ is scaled down to the UFM to obtain the set of possible UAV parameters $\left[T_p, T_d,\tau \right]$ on the UFM. The intersection of the two scaled manifolds gives the identified process parameters as illustrated in Fig. \ref{fig_pitch_alt_ident}. Identification results for both altitude and attitude dynamics are shown in Table \ref{tab:main_results}.



\begin{table}
    \centering
    \begin{tabular}{|c|c|c|c|c|c|c|c|}
    \hline  
         & & mean $T_d$ &  std $T_d$ & mean $\tau$ & std $\tau$\\\hline
        \multirow{4}{*}{UAV Design I} & Altitude & 2.669 & 0.6612 & 0.0464 & 0.0012\\\cline{2-6} 
        &Added 20 ms & 3.5077 & 1.203 & 0.0660   & 0.0041\\\cline{2-6}
          & Roll & 1.301 & 0.8395 & 0.0297 & 0.0042 \\ \cline{2-6}
          & Pitch & 1.6026 & 1.3094 & 0.0289 & 0.0066\\\hline 
         
        \multirow{4}{*}{UAV Design II} & Altitude & 1.8345 & 0.1754 & 0.0498  & 0.0017 \\\cline{2-6} 
         &Added 20 ms & 4.1755 &  0.5544&0.0636 & 0.0017 \\\cline{2-6}
          & Roll & 0.887 & 0.4574 & 0.0228 & 0.0053 \\ \cline{2-6}
          & Pitch & 0.6312 & 0.2874 & 0.0308 & 0.0071\\\hline 
    \end{tabular}
    \caption{Mean and standard deviation of the parameters identified. We get multiple steady state oscillations and hence we got multiple parameters measurements. The 20 ms delay is added in the altitude loop.}
    \label{tab:main_results}
    \vspace{-5mm}
\end{table}

The accuracy of the identified parameters in the identification experiments cannot be assessed because we cannot access the ground truth parameter values. We propose to
introduce some increment to the time delay in the flight controller software, which is known and, therefore can be compared against the estimated value. We chose to add a time delay of 20 ms to the altitude feedback loop. The manifolds' intersection is shown in Fig. \ref{fig:alt_identified} for the case of 20 ms additional time delay. Also Table \ref{tab:main_results} shows the identification results with the added delay. The experiemental video and results are given in \cite{paper_video}.

\vspace{-2mm}
\subsection{Comparison with the literature}
The identified parameters are compared to the parameters obtained by another state-of-the-art identification method which is the deep neural network with the modified relay feedback test (DNN-MRFT) method \cite{Ayyad2020RealTimeSI}. The parameters obtained by DNN-MRFT are close to the parameters obtained from our proposed identification methods for the altitude dynamics. However, DNN-MRFT failed to pick up the additional 20 ms introduced to the altitude dynamics. This might be attributed to the fact that DNN-MRFT performs identification by selecting from a discrete set of values, i.e. the  DNN used within this method is a classifier.


\begin{table}
    \centering
    \begin{tabular}{|c|c|c|c|c|}
    \hline
     & \multicolumn{2}{|c|}{Nominal dynamics} &\multicolumn{2}{|c|}{Additional 20 ms dynamics} \\\hline   
         &  Ours & DNN-MRFT & Ours & DNN-MRFT\\\hline
        $T_p$ & 0.0422 & 0.0321 &  
        0.0422 & 0.0321\\\hline
        $T_d$ & 2.669 & 1.6877 & 3.5077 & 1.6877\\\hline
        $\tau$ & 0.0464 & 0.06 & 0.066 & 0.06\\\hline
    \end{tabular}
    \caption{Comparison between the proposed identification method and DNN-MRFT. The identification was performed on the altitude loop of UAV Design I.}
    \label{tab:my_label}
    \vspace{-3mm}
\end{table}

\section{Conclusion}\label{sec_conclusion}
This paper presented a real-time system identification method utilizing the homogeneity properties of the MRFT and the  LPRS. The identification method presented can identify the drag coefficient and time delay of a UAV. The identification method is shown to be fast and stable and, therefore, applicable in real-time applications. 
The two MRFTs required for identification are performed within seconds and the parameters can be found within a fraction of a second, which facilitates the real-time applicability of the method.
The accuracy of the identification is demonstrated by showing the identification of the known incremental delay of the system. It was also shown that this method more accurately identifies the known incremental delays than another benchmark real-time identification method, which also provides very accurate controller tuning. 

So far, the altitude and the attitude dynamics of the UAV are considered for identification, but a similar identification could be employed in the future to identify the parameters of the underactuated lateral dynamics of the UAV as well. 




\bibliographystyle{ieeetr}
\bibliography{bare_jrnl}

\begin{thebibliography}{10}

\bibitem{ljung2010perspectives}
L.~Ljung, ``Perspectives on system identification,'' {\em Annual Reviews in
  Control}, vol.~34, no.~1, pp.~1--12, 2010.

\bibitem{Hann2009Sensors}
C.~E. Hann, I.~Singh-Levett, B.~L. Deam, J.~B. Mander, and J.~G. Chase,
  ``Real-time system identification of a nonlinear four-story steel frame
  structure—application to structural health monitoring,'' {\em IEEE Sensors
  Journal}, vol.~9, no.~11, pp.~1339--1346, 2009.

\bibitem{Qi2021Sensors}
C.~Qi, J.~Lin, Y.~Wu, and F.~Gao, ``A wiener model identification for creep and
  vibration linear and hysteresis nonlinear dynamics of piezoelectric
  actuator,'' {\em IEEE Sensors Journal}, vol.~21, no.~24, pp.~27570--27581,
  2021.

\bibitem{Jiang2022}
B.~Jiang, B.~Li, W.~Zhou, L.-Y. Lo, C.-K. Chen, and C.-Y. Wen, ``Neural network
  based model predictive control for a quadrotor {UAV},'' {\em Aerospace},
  vol.~9, p.~460, Aug. 2022.

\bibitem{Babaei2008Sensors}
F.~Hossein-Babaei and S.~M. Hosseini-Golgoo, ``Analyzing the responses of a
  thermally modulated gas sensor using a linear system identification technique
  for gas diagnosis,'' {\em IEEE Sensors Journal}, vol.~8, no.~11,
  pp.~1837--1847, 2008.

\bibitem{astrm1966_PE}
K.-J. {\AA}str\"{o}m and T.~Bohlin, ``Numerical identification of linear
  dynamic systems from normal operating records,'' in {\em Theory of
  Self-Adaptive Control Systems}, pp.~96--111, Springer {US}, 1966.

\bibitem{Hwang2022Sensors}
Y.~Hwang, Y.~Jeong, I.~S. Kweon, and S.~B. Choi, ``Identification of vehicle
  dynamics model and lever-arm for arbitrarily mounted motion sensor,'' {\em
  IEEE Sensors Journal}, vol.~22, no.~10, pp.~9843--9856, 2022.

\bibitem{Necmiye2019ACC}
S.~Oymak and N.~Ozay, ``Non-asymptotic identification of lti systems from a
  single trajectory,'' in {\em 2019 American Control Conference (ACC)},
  pp.~5655--5661, 2019.

\bibitem{Sarkar2022Finite}
T.~Sarkar, A.~Rakhlin, and M.~A. Dahleh, ``Finite time lti system
  identification,'' {\em J. Mach. Learn. Res.}, vol.~22, jul 2022.

\bibitem{Zheng2021multi}
Y.~Zheng and N.~Li, ``Non-asymptotic identification of linear dynamical systems
  using multiple trajectories,'' {\em IEEE Control Systems Letters}, vol.~5,
  no.~5, pp.~1693--1698, 2021.

\bibitem{astrm1984RFT}
K.~{\AA}str\"{o}m and T.~H\"{a}gglund, ``Automatic tuning of simple regulators
  with specifications on phase and amplitude margins,'' {\em Automatica},
  vol.~20, pp.~645--651, Sept. 1984.

\bibitem{Boiko2013}
I.~Boiko, ``Modified relay feedback test ({MRFT}) and tuning of {PID}
  controllers,'' in {\em Advances in Industrial Control}, pp.~25--79, Springer
  London, 2013.

\bibitem{Luyben1987}
W.~L. Luyben, ``Derivation of transfer functions for highly nonlinear
  distillation columns,'' {\em Industrial {\&} Engineering Chemistry Research},
  vol.~26, pp.~2490--2495, Dec. 1987.

\bibitem{Alfaro2021high}
V.~M. Alfaro and R.~Vilanova, ``Control of high-order processes: repeated-pole
  plus dead-time models' identification,'' {\em International Journal of
  Control}, vol.~0, no.~0, pp.~1--11, 2021.

\bibitem{Boiko2008Autotune}
I.~Boiko, ``Autotune identification via the locus of a perturbed relay system
  approach,'' {\em IEEE Transactions on Control Systems Technology}, vol.~16,
  no.~1, pp.~182--185, 2008.

\bibitem{Castellanos2008}
M.~I. Castellanos, I.~Boiko, and L.~Fridman, ``Parameter identification via
  modified twisting algorithm,'' {\em International Journal of Control},
  vol.~81, no.~5, pp.~788--796, 2008.

\bibitem{paper_video}
A.~Peringal, ``Experiment video.'' \url{https://youtu.be/LfBOwRgFaFs}, 2022.

\bibitem{multirotors}
A.~Ayyad, M.~Chehadeh, P.~H. Silva, M.~Wahbah, O.~A. Hay, I.~Boiko, and
  Y.~Zweiri, ``Multirotors from takeoff to real-time full identification using
  the modified relay feedback test and deep neural networks,'' {\em IEEE
  Transactions on Control Systems Technology}, vol.~30, no.~4, pp.~1561--1577,
  2022.

\bibitem{ayyad2021multirotors}
A.~Ayyad, M.~Chehadeh, P.~H. Silva, M.~Wahbah, O.~A. Hay, I.~Boiko, and
  Y.~Zweiri, ``Multirotors from takeoff to real-time full identification using
  the modified relay feedback test and deep neural networks,'' {\em IEEE
  Transactions on Control Systems Technology}, 2021.

\bibitem{BOIKO2005677}
I.~Boiko, ``Oscillations and transfer properties of relay servo systems—the
  locus of a perturbed relay system approach,'' {\em Automatica}, vol.~41,
  no.~4, pp.~677--683, 2005.

\bibitem{boiko2007analysis}
I.~Boiko, L.~Fridman, A.~Pisano, and E.~Usai, ``Analysis of chattering in
  systems with second-order sliding modes,'' {\em IEEE transactions on
  Automatic control}, vol.~52, no.~11, pp.~2085--2102, 2007.

\bibitem{Ayyad2020RealTimeSI}
A.~Ayyad, M.~Chehadeh, M.~I. Awad, and Y.~H. Zweiri, ``Real-time system
  identification using deep learning for linear processes with application to
  unmanned aerial vehicles,'' {\em IEEE Access}, vol.~8, pp.~122539--122553,
  2020.

\end{thebibliography}
\vspace{-10mm}

\begin{IEEEbiography}[{\includegraphics[width=1in,height=1.25in,clip,keepaspectratio]{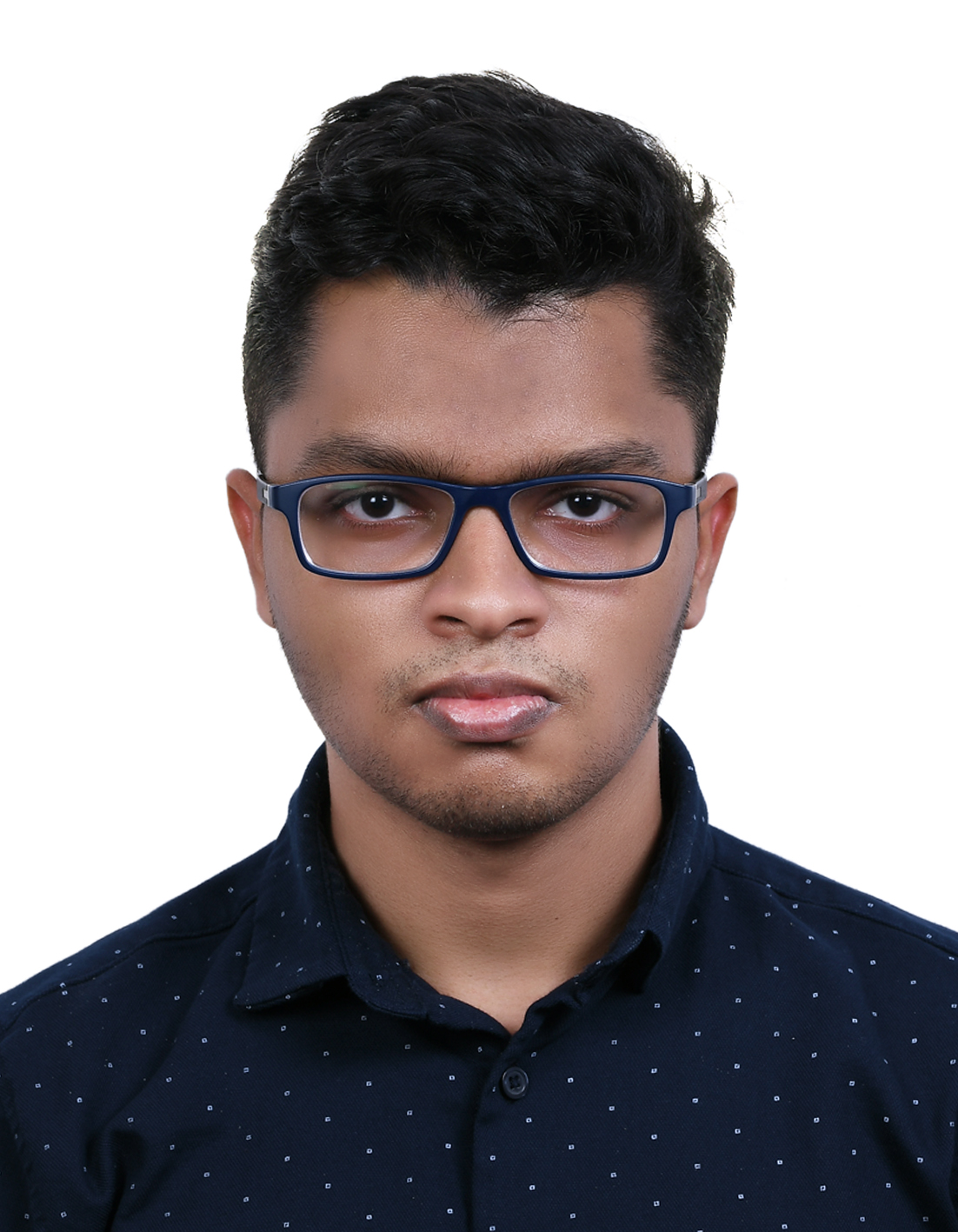}}]{Anees Peringal} received his BSc. degree in Aerospace engineering from Khalifa University, Abu Dhabi. He is currently pursuing an MSc. degree in Aerospace engineering at Khalifa University. He is interested in research related to control of dynamic systems and autonomous robotics. 

\end{IEEEbiography}
\vspace{-10mm}

\begin{IEEEbiography}[{\includegraphics[width=1in,height=1.25in,clip,keepaspectratio]{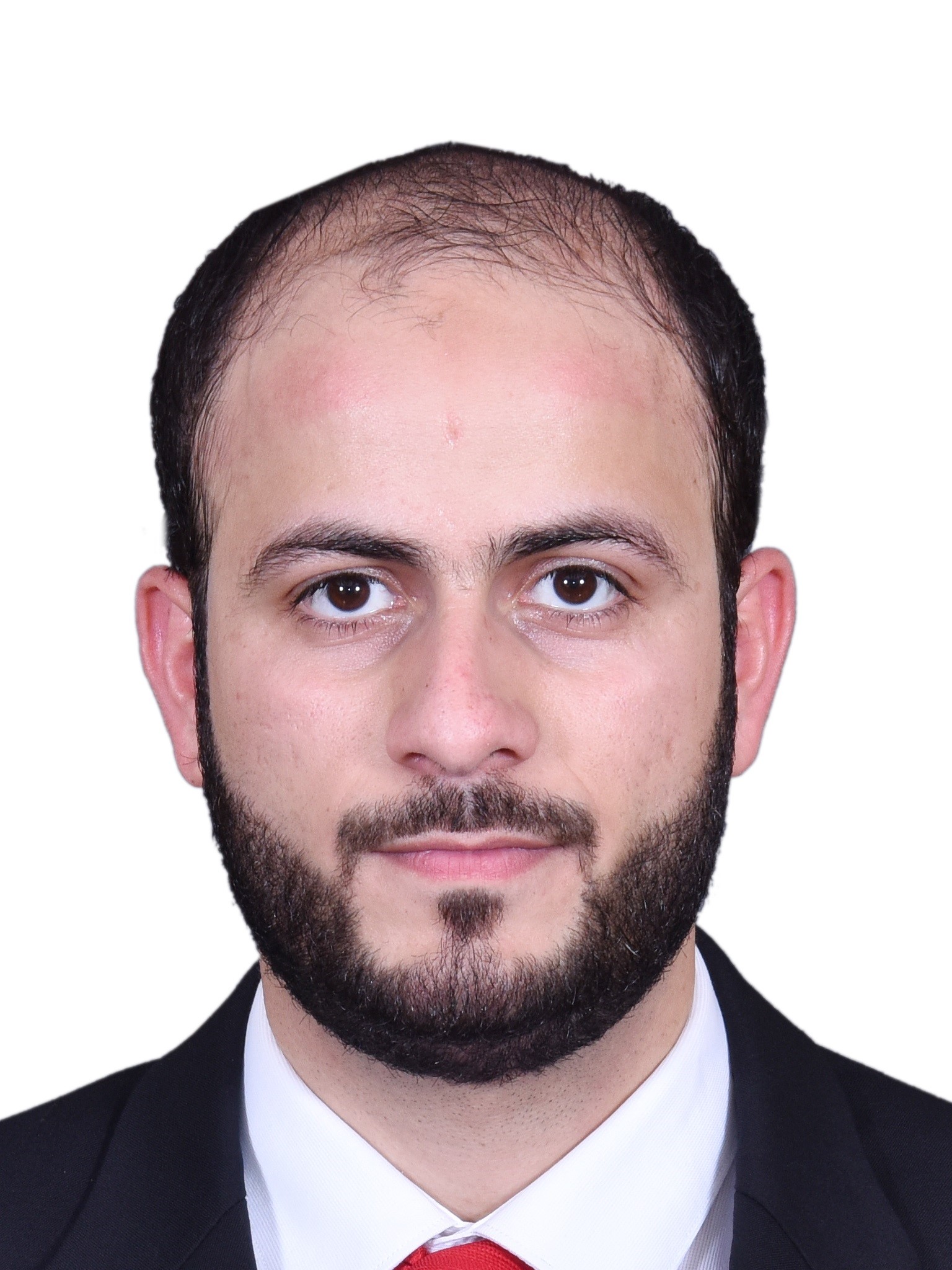}}]{Mohamad Chehadeh} received his MSc. in Electrical Engineering from Khalifa University, Abu Dhabi, UAE, in 2017. He is currently with Khalifa University Center for Autonomous Robotic Systems (KUCARS). His research interest is mainly focused on identification, perception, and control of complex dynamical systems utilizing the recent advancements in the field of AI.
\end{IEEEbiography}
\vspace{-8mm}
\begin{IEEEbiography}[{\includegraphics[width=1in,height=1.25in,clip,keepaspectratio]{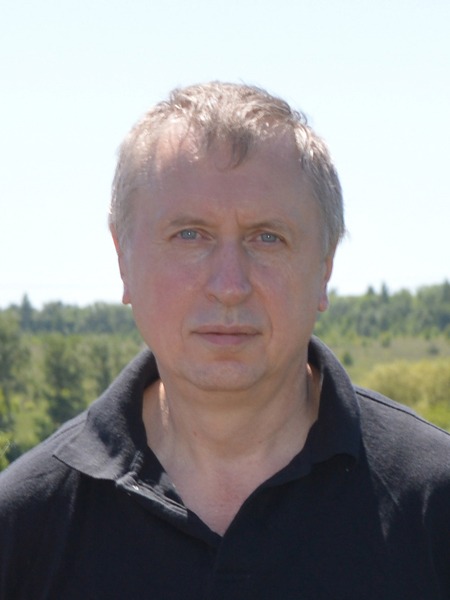}}]{Igor Boiko} received his MSc, PhD and DSc degrees from Tula State University and Higher Attestation Commission, Russia. His research interests include frequency-domain methods of analysis and design of nonlinear systems, discontinuous and sliding mode control systems, PID control, process control theory and applications. Currently, he is a Professor with Khalifa University, Abu Dhabi, UAE.
\end{IEEEbiography}

\begin{IEEEbiography}[{\includegraphics[width=1in,height=1.25in,clip,keepaspectratio]{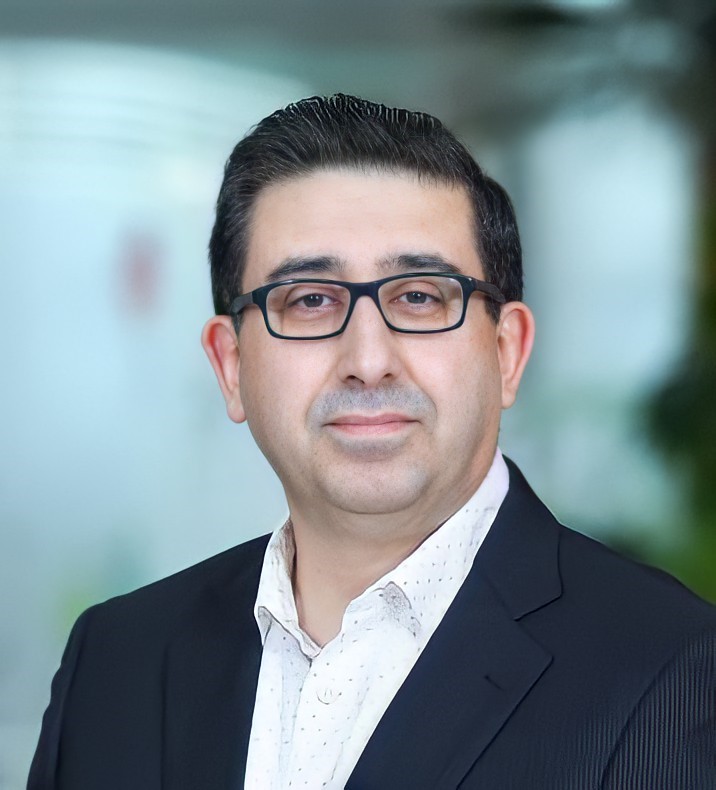}}]{Yahya Zweiri} (Member, IEEE) received the Ph.D. degree from the King’s College London in 2003. He is currently an Associate Professor with the Department of Aerospace Engineering and director of the Advanced Research and Innovation Center - Khalifa University, United Arab Emirates. He was involved in defense and security research projects in the last 20 years at the Defense Science and Technology Laboratory, King’s College London, and the King Abdullah II Design and Development Bureau, Jordan. He has published over 130 refereed journals and conference papers and filed ten patents in USA and UK. His main research is in the area of robotic systems for extreme conditions with particular emphasis on applied AI aspects and neuromorphic vision systems.
\end{IEEEbiography}

\end{document}